\documentstyle[12pt]{article}

\author{Hao Chen\\
Department of Mathematics\\
Zhongshan University\\
Guangzhou,Guangdong 510275\\
People's Republic of China\\
and\\
Department of Computer Science\\
National University of Singapore\\
Singapore 117543\\
Republic of Singapore}
\title{New Invariants and Separability Criterion of the Mixed States: Multipartite Case}
\date{July,2001}

\begin{document}

\maketitle
\begin{abstract}
We introduce algebraic sets in the products of the complex projective spaces for the mixed states in a multipartite quantum systems as their invariants under local unitary operations. The algebraic sets have to be the union of the  linear subspaces if the mixed state is separable, and thus we give a new criterion of separability for the mixed states in multipartite quantum systems. Some examples are studied by  our criterion. Our invariants also can be used to distinguish inequivalent mixed states under local unitary operations.

\end{abstract}

In recent years it became clear that entanglement is one of the most important ingredients and resources of quantum infromation processing(see [1],[2]), and thus stimulated tremendous studies of quantum entanglements of both bipartite and multipartite systems, for a survey we refer to [3],[8] and [9]. For multipartite case, the criterion of Peres-Horodecki said that a separable mixed state must necessarily have positive partial transposes under all cuts of the whole system(PPT). It is also observed that the mixed states with PPT cannot be distilled (bound entanglement). In [14], the separability of the mixed states in mutipartite systems was studied by linear maps. In $2 \times 2 \times 2$ system, an example of rank 4 entangled mixed state  with PPT is presented in the context of unextendible product bases (UPB,[10]), and this state is actually separable under the  A:BC, AB:C and AC:B cuts. For more examples related to UPB, we refer to [10] and [11]. For ``low''  rank PPT  mixed states in $2 \times 2 \times N$ system ,it is proved that they have to be separable([13]). Examples of 3-party and 4-party entangled mixed states with very interesting properties were also given and studied in [12] and [15].\\

In our previous work [20] we introduced  algebraic sets (ie., the zero locus of several multi-variable homogeneous polynomials, see [18]) for the mixed states in bipartite quantum system and proved that 1)the algebraic-geometric and metric  (Hermitian geometric)  properties of these algebraic sets are invariant when local unitary operations are applied to the mixed states, and thus any algebraic-geometric or Hermitian geometric invariant of the algebraic set is an invariant of the corresponding mixed state; 2)The algebraic sets have to be linear if the mixed state is separable. This established a connection between Quantum Entanglement and both  Algebraic and Hermitian Geometry.\\

In this letter, we continue our study for multipartite quantum systems. Basically we  introduce algebraic sets  in the product of the complex projective  spaces (see [18]) for any given mixed state in a mutlipartite quantum system with the following two properties:\\

1) When we apply local unitary operations to the mixed state the corresponding  algebraic sets are  changed by a linear transformation, and thus these invariants can be used to distinguish inequivalent mixed states under local unitary operations;\\

2)The algebraic sets are  linear (the union of some linear subspaces) if the mixed state is separable, and thus we give a new separability criterion.\\

For the algebraic geometry used in this paper, we refer to the nice book [18].\\

We introduce the algebraic sets of the mixed states and prove the results for tripartite case.  Then the multipartite case is similar and we just generalize directly.\\

Let $H=C_{A}^m \otimes C_{B}^n \otimes C_{C}^l$ and the standard orthogonal base is $|ijk>$, where, $i=1,...,m$,$j=1,...,n$ and $k=1,...,l$, and $\rho$ is a mixed state on $H$. We represent the matrix of $\rho$ in the base $\{|111>,...|11l>,...,|mn1>,...,|mnl>\}$ as
 $\rho=(\rho_{ij,i'j'})_{1 \leq i ,i' \leq m, 1 \leq j ,j' \leq n}$, and  
$\rho_{ij,i'j'}$ is a $l \times l$ matrix. Consider $H$ as a bipartite system as $H=(C_{A}^m \otimes C_{B}^n) \otimes C_{C}^l$, then we have $V_{AB}(\rho)=\{(r_{11},...,r_{mn}) \in C^{mn}:det( \Sigma r_{ij}r_{i'j'}^{*} \rho_{ij,i'j'})=0\}$ defined as in [20]. When the finer cut A:B:C is considered we define $V_{AB}^{A:B}(\rho)$ as follows.\\

{\bf Definition 1.}{\em  Let $\phi:CP^{m-1} \times CP^{n-1} \rightarrow CP^{mn-1}$  be the mapping defined by\\

$$
\begin{array}{ccccccccc}
\phi(r_1^1,...r_m^1,r_1^2,...,r_n^2)=(r_1^1r_1^2,...,r_i^1r_j^2,...r_m^1r_n^2)
\end{array}
(1)
$$

(ie., $r_{ij}=r_i^1 r_j^2$ is introduced.)

Then $V_{AB}^{A:B}(\rho)$ is defined as the preimage $\phi^{-1}(V_{AB}(\rho))$.}\\

Similarly $V_{BC}^{B:C},V_{AC}^{A:C}$ can be defined. In the following statement we just state the result for $V_{AB}^{A:B}$. The conclusion holds similarly for other $V's$.\\

From this definition and Theorem 2 in [20] we immediately have the following result.\\

{\bf Theorem 1.} {\em $V_{AB}^{A:B}(\rho)$ is an algebraic set in $CP^{m-1} \times CP^{n-1}$.}\\

{\bf Theorem 2.}{\em  Let $T=U_{A} \otimes U_{B} \otimes U_{C}$, where $U_{A},U_{B}$ and $U_{C}$ are the local operations (ie., unitary linear transformation) on $C_{A}^m, C_{B}^n$ and $C^l$ rescpectively. Then $V_{AB}^{A:B}(T(\rho))=U_{A}^{-1} \times U_{B}^{-1}(V_{AB}^{A:B}(\rho))$,  that is $V_{AB}^{A:B}(\rho)$ is a ``invariant'' upto a linear transformation of $CP^{m-1} \times CP^{n-1}$  of the mixed state $\rho$ under local unitary operations.}\\

{\bf Proof.} Let $U_{A}=(u_{ij}^{A})_{1 \leq i \leq m,1 \leq j \leq m}$, $U_{B}=(u_{ij}^{B})_{1 \leq i \leq n, 1 \leq j \leq n}$ and $U_{C}=(u_{ij}^{C})_{1 \leq i \leq l, 1 \leq j \leq l}$, be the matrix in the standard orthogonal bases.\\

Recall the proof of Theorem 1 in [20], we have $V_{AB}(T(\rho))=(U_{A} \otimes U_{B})^{-1}(V_{AB}(\rho))$ under the coordinate change\\

$$
\begin{array}{cccccccc} 
r_{kw}'=\Sigma_{ij} r_{ij} u_{ik}^{A} u_{jw}^{B}\\
=\Sigma_{ij}r_i^1 r_j^2 u_{ik}^{A} u_{jw}^{B}\\
=\Sigma_{ij} (r_{i}^1 u_{ik}^A) (r_{j}^2 u_{jw}^B)\\
=(\Sigma_i r_i^1 u_{ik}^A)(\Sigma_j r_j^2 u_{jw}^B)
\end{array}
(2)
$$

for $k=1,...,m,w=1,...,n$. Thus our conclusion follows from the definition.\\

{\bf Remark 1.} Since $U_{A}^{-1} \times U_{B}^{-1}$ certainly preserves the  (product) Fubini-Study metric of $CP^{m-1} \times CP^{n-1}$, we know that all metric properties of $V_{AB}^{A:B}(\rho)$ are preserved when the local unitary operations are applied to the mixed state $\rho$.\\

In the following statement we give the separability criterion of the mixed state $\rho$ under the cut A:B:C. The ``linear subspace of $CP^{m-1} \times CP^{n-1}$'' means the product of a linear subspace in $CP^{m-1}$ and a linear subspace in $CP^{n-1}$.\\

{\bf Theorem 3.}{\em If $\rho$ is a separable mixed state on $H=C_{A}^m \otimes C_{B}^{n} \otimes C_{C}^l$ under the cut A:B:C, $V_{AB}^{A:B}(\rho)$ is a linear subset of $CP^{m-1} \times CP^{n-1}$, ie., it is the union of the linear subspaces.}\\

{\bf Proof.} We first consider the separability of $\rho$ under the cut AB:C,ie., $\rho= \Sigma_{f=1}^g p_f P_{a_f \otimes b_f}$, where $a_f \in C_{A}^m \otimes C_{B}^n$ and $b_f \in C_{C}^l$ for $f=1,...,g$. Consider the separability of $\rho$ under the cut A:B:C, we have  $a_f=a_f' \otimes a_f''$ , $a_f' \in C_{A}^m, a_f'' \in C_{B}^m$. Let $a_f=(a_f^1,...,a_f^{mn}), a_f'=(a_f'^1,...,a_f'^m)$ and $a_f''(a_f''^1,...,a_f''^n)$ be the coordinate forms with the standard orthogonal basis $\{|ij>\}$, $\{|i>\}$ and $\{|j>\}$ respectively, we have that $a_f^{ij}=a_f'^i a_f''^j$. Recall the proof of Theorem 3 in [20], the diagonal entries of $G$ in the proof of Theorem 3 in [20] are\\

$$
\begin{array}{cccccccc}
\Sigma_{ij}r_{ij} a_f^{ij}=\\
\Sigma_{ij} r_i^1 a_f'^i r_j^2 a_f''^j=\\
(\Sigma_i r_i^1 a_f'^i)(\Sigma_j r_j^1 a_f''^j)
\end{array}
(3)
$$

Thus as argued in the proof of Theorem 3 of [20], $V_{AB}^{A:B}(\rho)$ has to be the zero locus of the multiplications of the linear forms in (3). The conclusion is proved.\\

For the mixed state $\rho$ in the multipartite system $H=C_{A_1}^{m_1} \otimes \cdots \otimes C_{A_k}^{m_k}$, we want to study the entanglement under the cut $ A_{i_1}:A_{i_2}:...:A_{i_l}:(A_{j_1}...A_{j_{k-l}})$, where $\{i_1,...,i_l\} \cup \{j_1,...j_{k-l}\}=\{1,...k\}$. We can define the set $V_{A_{i_1}...A_{i_l}}^{A_{i_1}:...:A_{i_l}}(\rho)$ similarly. We have the following results.\\

{\bf Theorem 1'.} {\em $V_{A_{i_1}...A_{i_l}}^{A_{i_1}:...:A_{i_l}}(\rho)$ is an algebraic set in in $CP^{m_{i_1}-1} \times CP^{m_{i_l}-1}$.}\\

{\bf Theorem 2'.}{\em  Let $T=U_{A_{i_1}} \otimes \cdots \otimes U_{A_{i_l}} \otimes U_{j_1...j_{k-l}}$, where $U_{A_{i_1}},...,U_{A_{i_l}}, U_{j_1...j_{k-l}}$ are the local operations (ie., unitary linear transformation) on $C_{A_{i_1}}^{m_{i_1}},...,C_{A_{i_l}}^{m_{i_l}}$  and $(C_{A_{j_1}}^{m_{j_1}} \otimes ...\otimes C_{A_{j_{k-l}}}^{m_{j_{k-l}}})$ rescpectively. Then $V_{A_{i_1}...A_{i_l}}^{A_{i_1}:...:A_{i_l}}(T(\rho))=U_{A_{i_1}}^{-1} \times \cdots \times U_{A_{i_l}}^{-1}(V_{A_{i_1}...A_{i_l}}^{A_{i_1}:...:A_{i_l}}(\rho))$.}\\

{\bf Theorem 3'.}{\em If $\rho$ is a separable mixed state on $H=C_{A_1}^{m_1} \otimes \cdots \otimes C_{A_k}^{m_k}$ under the cut $ A_{i_1}:A_{i_2}:...:A_{i_l}:(A_{j_1}...A_{j_{k-l}})$, $V_{A_{i_1}...A_{i_l}}^{A_{i_1}:...:A_{i_l}}$ is a linear subset of $CP^{m_{i_1}-1}  \times ... \times CP^{m_{i_l}-1}$,ie., it is the union of the linear subspaces.}\\

In the following we study and give some examples of mixed states based on our above results.\\

{\bf Example 1 (J.Smolin [15]).} Let $H=C_{A}^2 \otimes C_{B}^2 \otimes C_{C}^2 \otimes C_{D}^2$ and \\

$$
\begin{array}{cccccccc}
|\psi^{\pm}>= \frac{1}{\sqrt {2}}(|01>+|10>)\\
|\phi^{\pm}>= \frac{1}{\sqrt {2}}(|00>+|11>)\\
\rho= \frac{1}{4}(P_{|\phi^{+}>^{AB}} \otimes P_{|\phi^{+}>^{CD}} +P_{|\phi^{-}>^{AB}} \otimes P_{|\phi^{-}>^{CD}} +\\
 P_{|\psi^{+}>^{AB}} \otimes P_{|\psi^{+}>^{CD}}+ P_{|\psi^{-}>^{AB}} \otimes P_{|\psi^{-}>^{CD}})
\end{array}
(4)
$$

This is a rank 4 mixed state. We can calculate its ``invariants'' $V_{AB}^{A:B}$ easily as 

$$
\begin{array}{cccc}
V_{AB}^{A:B:C:D}=\{(r_0^1,r_1^1,r_0^2,r_1^2) \in CP^1 \times CP^1 :\\
(r_0^1 r_0^2 + r_1^1 r_1^2)(r_0^1 r_0^2 - r_1^1 r_1^2)(r_0^1 r_1^2 + r_1^1 r_0^2)(r_0^1 r_1^2 - r_1^1 r_0^2)=0\}
\end{array}
(5)
$$

Thus $V_{AB}^{A:B}$ is the union of the four algebraic varieties (corresponding to 4 terms in (5)). Each variety, e.g., $\{(r_0^1,r_1^1,r_0^2,r_1^2) \in CP^1 \times CP^1: r_0^1 r_0^2 + r_1^1 r_1^2=0\}$, is not linear,  since the degree 2 polynomial $r_0^1 r_0^2 + r_1^1 r_1^2$ cannot be factorize to 2 linear forms. Thus $\rho$ is entangled under the cut A:B:C:D. However from a result in [9], it is separable under the cuts AB:CD,AC:BD,AD:BC.\\

The following example can be thought as a generalization of Smolin's mixed state.\\

{\bf Example 2.} Let $H=C_{A}^2 \otimes C_{B}^2 \otimes C_{C}^2 \otimes C_{D}^2$ and $h_1,h_2,h_3,h_4$ (understood as row vectors)are 4 mutually orthogonal unit vectors in $C^4$. Consider the $16 \times 4$ matrix $T$ with 16 rows as\\
 $T=(a_1h_1^{\tau},0,0,a_2 h_2^{\tau},0, a_3 h_3^{\tau},a_4 h_4^{\tau},0,0, a_5 h_3^{\tau}, a_6 h_4^{\tau},0, a_7 h_1^{\tau},0,0,a_8 h_2^{\tau})^{\tau}$. Let \\ $\phi'_1,\phi'_2,\phi'_3,\phi'_4$ be 4 vectors in $H$ whose expansions with the base $|0000>,|0001>,|0010>,|0011>,|0100>,|0101>,|0110>,|0111>,|1000>$,\\$|1001>,|1010>,|1011>,|1100>,|1101>,|1110>,|1111> $ are exactly the 4 columns of the matrix $T$ and $\phi_1,\phi_2,\phi_3,\phi_4$ are the normalized unit vectors of $\phi'_1,\phi'_2, \phi'_3, \phi'_4$. Let $\rho=\frac{1}{4}(P_{\phi_1} +P_{\phi_2} +P_{\phi_3} +P_{\phi_4})$.\\

It is easy to check that when $h_1=(1,1,0,0),h_2=(1,-1,0,0), h_3=(0,0,1,1), h_4=(0,0,1,-1)$ and $a_1=a_2=a_3=a_4=1$. It is just the Somolin's example in [15]\\

Now we prove that $\rho$ is invariant under the partial transposes of the cuts AB:CD,AC:BD,AD:BC.\\

Let  the ``representation'' matrix $T=(b_{ijkl})_{i=0,1,j=0,1,k=0,1,l=0,1}$ is the matrix with columns corresponding the expansions of $\phi_1,\phi_2,\phi_3,\phi_4$.Then we can consider that $T=(T_1,T_2,T_3,T_4)^{\tau}$ is blocked matrix of size $4 \times 1$ with each block $T_{ij}=(b_{kl })_{k=0,1,l=0,1}$ a $4 \times 4$ matrix,where $ij=00,01,10,11$. Because $h_1,h_2,h_3,h_4$ are mutually orthogonal unit vectors we can easily check that $T_{ij} (T_{i'j'}^{*})^{\tau}=T_{i'j'} (T_{ij}^{*})^{\tau}$ Thus it is invariant when the partial transpose of the cut AB:CD is applied.\\

With the same methods we can check that $\rho$ is invariant when the partial transposes of the cuts AC:BD, AD:BC  are applied. Hence $\rho$ is PPT under the cuts AB:CD, AC:BD,AD:BC. Thus from a result in [9] we know $\rho$ is separable under these cuts AB:CD, AC:BD,AD:BC.\\

Now we want to prove $\rho$ is entangled under the cut A:BCD by computing $V_{BCD}(\rho)$. From the arguments in [20] and this paper, we can check that $V_{BCD}(\rho)$ is the locus of the condition: $a_1 h_1 r_{000} + a_2 h_2 r_{011} +a_3 h_3 r_{101} +a_4 h_4 r_{110}$ and $a_7 h_1 r_{100} + a_8 h_2 r_{111} +a_5 h_3 r_{001} +a_6 h_4 r_{010}$ are linear dependent. This is equivalent to the condition that the matrix (6) is of  rank 1.\\

$$
\left(
\begin{array}{cccccc}
a_7 r_{100} & a_8 r_{111} & a_5 r_{001} & a_6 r_{010}\\
a_1 r_{000} & a_2 r_{011} & a_3 r_{101} & a_4 r_{110}
\end{array}
\right)
(6)
$$

From [18] pp. 25-26 we can check that $V_{BCD}(\rho)$ is exactly the famous Segre variety in algebraic geometry. It is irreducible and thus cannot be linear. From Theorem 3 in [20], $\rho$ is entangled under the cut A:BCD. Similarly we can prove that $\rho$ is entangled under the cuts B:ACD, C:ABD, D:ABC.\\

Now we compute $V_{AB}^{A:B}$. From the arguments in [20] and Definition 1 , it is just the locus of the condition that the vectors $h_1(a_1 r_0^1 r_0^2 +a_7 r_1^1 r_1^2)$, $h_3 (a_3 r_0^1 r_1^2 +a_5 r_1^1 r_0^2)$, $h_4(a_4 r_0^1 r_1^2 +a_6  r_1^1 r_0^2)$, $h_2 (a_2 r_0^1 r_0^2 +a_8 r_1^1 r_1^2)$ are linear dependent. Since $h_1,h_2,h_3,h_4$ are mutually orthogonal unit vectors,\\

$$
\begin{array}{cccccccccc}
V_{AB}^{A:B}=\{(r_0^1,r_1^1,r_0^2,r_1^2) \in CP^1 \times CP^1:\\
(a_1 r_0^1 r_0^2 +a_7 r_1^1 r_1^2)(a_3 r_0^1 r_1^2 +a_5 r_1^1 r_0^2)(a_4 r_0^1 r_1^2 +a_6  r_1^1 r_0^2)(a_2 r_0^1 r_0^2 +a_8 r_1^1 r_1^2)=0\}
\end{array}
(7)
$$

Let $\lambda_1=-a_1/a_7,\lambda_2=-a_3/a_5, \lambda_3=-a_4/a_6, \lambda_4=-a_2/a_8$ and consider the family of the mixed states $\{\rho_{\lambda_{1,2,3,4}}\}$, we want to prove the following statement.\\

{\bf Theorem 4.} {\em There are uncountably many members which are inequivalent
under the local operations on $H=C_{A}^2 \otimes C_{B}^2 \otimes C_{C}^2 \otimes C_{D}^2$ in the above family of mixed states on $H$.}\\

{\bf Proof.} From the above computation, $V_{AB}^{A:B}(\rho_{\lambda_{1,2,3,4}})$ is the union of the following 4 algbraic varieties in $CP^1 \times CP^1$.\\

$$
\begin{array}{ccccccccccc}
V_1=\{(r_0^1,r_1^1,r_0^2,r_1^2) \in CP^1 \times CP^1:r_0^1 r_0^2 - \lambda_1 r_1^1 r_1^2=0\}\\
V_2=\{(r_0^1,r_1^1,r_0^2,r_1^2) \in CP^1 \times CP^1:r_0^1 r_1^2 - \lambda_2 r_1^1 r_0^2=0\}\\
V_3=\{(r_0^1,r_1^1,r_0^2,r_1^2) \in CP^1 \times CP^1:r_0^1 r_1^2 - \lambda_3 r_1^1 r_0^2=0\}\\
V_4=\{(r_0^1,r_1^1,r_0^2,r_1^2) \in CP^1 \times CP^1:r_0^1 r_0^2 - \lambda_4 r_1^1 r_1^2=0\}
\end{array}
(8)
$$

From Theorem 2, if $\rho_{\lambda_{1,2,3,4}}$ and $\rho_{\lambda'_{1,2,3,4}}$ are equivalent by a local operation, there must exist 2 fractional linear transformations $T_1, T_2$ of $CP^1$ such that $T=T_1 \times T_2$ (acting on  $CP^1 \times CP^1$) transforms the 4 varieties $V_1,V_2,V_3,V_4$  of $\rho_{\lambda_{1,2,3,4}}$ to the  4 varieties $V'_1,V'_2,V'_3,V'_4$  of $\rho_{\lambda'_{1,2,3,4}}$,ie., $T(V_i)=V'_j$.\\

Introduce the inhomogeneous coordinates $x_1=r_0^1/r_1^1,x_2=r_0^2/h_1^2$. Let $T_1(x_1)=(ax_1+b)/(cx_1+d)$.  Suppose $T(V_i)=V'_i, i=1,2,3,4$. Then we have $ab \lambda_1= cd \lambda'_1 \lambda'_2$ and $ab \lambda_4=cd \lambda'_3 \lambda'_4$. Hence $\lambda_1 \lambda'_3 \lambda'_4 =\lambda'_1 \lambda'_2 \lambda_4$. This means that there are some algebraic relations of parameters if the $T$ exists. Similarly we can get the same conclusion for the other possibilities $T(V_i)=V'_j$. This implies that there are some algebraic relations of parameters $\lambda_{1,2,3,4}$ and $\lambda'_{1,2,3,4}$ if $\rho_{\lambda_{1,2,3,4}}$ and $\rho_{\lambda'_{1,2,3,4}}$ are  equivalent  by a local operation. Hence our conclusion follows immediately.\\

The author acknowledges the support from NNSF China, Information Science Division, grant 69972049.\\

e-mail: dcschenh@nus.edu.sg\\

\begin{center}
REFERENCES
\end{center}

1. C.H. Bennett and P.W.Shor, Quantum Information Theory, IEEE Trans. Inform. Theory, vol.44(1998),Sep.\\

2.J.Preskill, Physics 229:Advanced Mathematical Methodes of Physics--Quantum Computation and Information (California Institute of Technology, Pasadena, CA,1998), http://www.theory.caltech.edu/preskill/ph229/.\\

3.B.M.Terhal, Detecting quantum entanglements, quant-ph/0101032\\

4.R.Werner, Phys.Rev., A40,4277(1989)\\

5.A.Peres, Phys.Rev. Lett. 77,1413(1996)\\

6.M.Horodecki, P.Horodecki and R.Horodecki, Phys. Lett. A 223,8 (1996)\\

7.P.Horodecki, Phys.Lett. A 232 333(1997)\\

8.M.Horodecki , P.Horodecki and R. Horodecki, in ``Quantum Infomation---Basic concepts and experiments'', Eds, G.Alber and M.Wiener (Springer Berlin)\\

9.M. Lewenstein, D Bruss, J.I.Cirac, B.Krus, M.Kus, J. Samsonowitz , A.Sanpera and R.Tarrach, J.Mod. Optics, 47, 2481,(2000), quantu-ph/0006064\\

10.C.H.Bennett, D.P.DiVincenzo, T.Mor, P.W. Shor, J.A.Smolin and T.M. Terhal, Phys.Rev. Lett., 82: 5385-5388,1999\\

11.D.P.DiVincenzo, T.Mor, P.W.Shor, J.A.Smolin and B.M.Terhal, to appear in Comm.Math. Phys, quant-ph/9908070 \\

12.W.Dur,J.I.Cirac and R.Tarrach, Phys.Rev.Lett. 83,3562(1999), quant-ph/9903018\\

13.S.Karnas and M.Lewenstein, quant-ph/0102115\\

14.P.Horodecki, quant-ph/0006071\\

15.J.A.Smolin, quant-ph/0001001\\

16 N.Linden and P.Popescu, Fortsch. Phys.,46, 567(1998), quant-ph/9711016\\

17.B.Terhal and P.Horodecki, Phys. Rev. A61, 040301(2000); A.Sanpera, D.Bruss and M.Lewenstein, Phys. Rev. A63 R03105(2001)\\

18.J.Harris, Algebraic geometry, A first Course, Gradute Texts in Mathematics, 133, Springer-Verlag, 1992,
  especially its Lecture 9  ``Determinantal Varieties''\\

19.E.Arbarello, M.Cornalba, P.A.Griffiths and J.Harris, Geometry of algebraic curves,Volume I, Springer-Verlag, 1985, Chapter II''Determinantal Varieties''\\

20.Hao Chen, New invariants and separability criterion of the mixed states: bipartite case, preprint July,2001\\

21.Hao Chen, Families of inequivalent entangled mixed states related to Calabi-Yau manifolds, in preparation\\

22.Hao Chen, Quantum entanglement and geometry of determinantal varieties, in preparation\\

23.Hao Chen, Geometry of determinantal varieties associated with unextendible product bases and bound entanglement, in preparation\\

\end{document}